\documentclass{article}

\usepackage[preprint]{neurips_2025}
\usepackage[utf8]{inputenc} % allow utf-8 input 
\usepackage[T1]{fontenc}    % use 8-bit T1 fonts
\usepackage{hyperref}       % hyperlinks
\usepackage{url}            % simple URL typesetting
\usepackage{booktabs}       % professional-quality tables
\usepackage{amsfonts}       % blackboard math symbols
\usepackage{nicefrac}       % compact symbols for 1/2, etc.
\usepackage{microtype}      % microtypography
\usepackage{xcolor}         % colors
\usepackage{graphicx}
\usepackage{wrapfig}  % 在导言区添加

\usepackage{authblk}
\usepackage{subfiles}

\usepackage{algorithm}
\usepackage[noend]{algpseudocode}

\usepackage{color}

\usepackage{hyperref} 
\hypersetup{
    colorlinks=true,
    linkcolor=red,
    urlcolor=blue,
    linktoc=all,
    citecolor=cyan
}

\usepackage{enumitem}
\usepackage{lipsum}
\setlist[itemize]{leftmargin=*}
 
\definecolor{BgWhite}{rgb}{1,1,1} 
\definecolor{Gray}{rgb}{0.5,0.5,0.5} 
\definecolor{mygray}{gray}{0.6}

\usepackage{multirow}
\usepackage[utf8]{inputenc} % allow utf-8 input 
\usepackage{amsmath}

\usepackage{xcolor}

\title{ Auto-Regressive Surface Cutting}
\author{
Yang Li$^{1,*}$, Victor Cheung$^{1,*}$, Xinhai Liu$^1$, Yuguang Chen$^{2,\dagger}$, Zhongjin Luo$^{3,\dagger}$,  \\ \vspace{0.01cm} 
Biwen Lei$^1$, Haohan Weng$^{4,\dagger}$, Zibo Zhao$^{1,5}$, Jingwei Huang$^1$, Zhuo Chen$^1$, Chunchao Guo$^{1,\ddag}$ \\ \vspace{0.3cm}
$^1$ Tencent Hunyuan, $^2$SYSU, $^3$CUHKSZ, $^4$SCUT, $^5$ShanghaiTech \\ \vspace{0.3cm}
\url{https://victorcheung12.github.io/seamgpt}
}

\begin{document}

\maketitle

\begin{abstract}
 
Surface cutting is a fundamental task in computer graphics, with applications in UV parameterization, texture mapping, and mesh decomposition. However, existing methods often produce technically valid but overly fragmented atlases that lack semantic coherence.
We introduce SeamGPT, an auto-regressive model that generates cutting seams by mimicking professional workflows.
Our key technical innovation lies in formulating surface cutting as a next token prediction task: sample point clouds on mesh vertices and edges, encode them as shape conditions, and employ a GPT-style transformer to sequentially predict seam segments with quantized 3D coordinates.
Our approach achieves exceptional performance on UV unwrapping benchmarks containing both manifold and non-manifold meshes, including artist-created, and 3D-scanned models. 
In addition, it enhances existing 3D segmentation tools by providing clean boundaries for part decomposition.
\end{abstract}

\begin{figure}[!ht]
    \centering
    \includegraphics[width=1\linewidth]{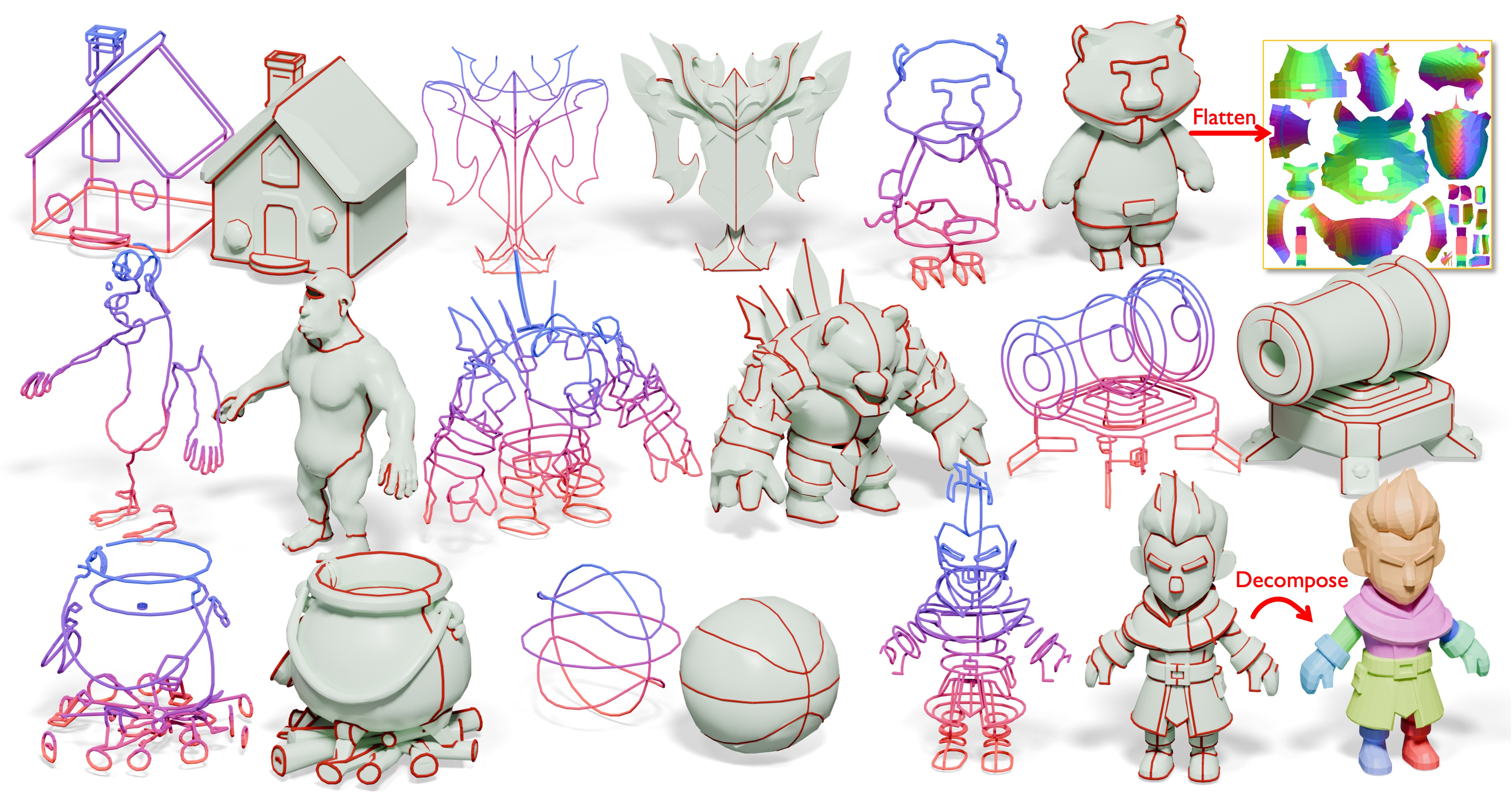}
    \caption{SeamGPT generates surfaces cutting seams, facilitating UV flatten and part decomposition.}
    \label{fig:enter-label}
\end{figure}

% \newpage
\section{Introduction}
\label{sec:intro}

% what is it
Surface cutting – the process of decomposing 3D mesh geometry through topological incisions – 
serves as a fundamental operation for numerous applications, including UV parameterization, texture mapping, and digital fabrication. 
This technique also enables critical downstream tasks such as remeshing, part-based editing, and shape decomposition.

The primary objective of surface cutting is to achieve low-distortion surface flattening. 
Current industry-standard tools (e.g., XAtlas or Blender's Smart UV Project) optimize for minimal distortion when flattening meshes into 2D atlases.
While these methods produce technically valid cuts, they often generate over-fragmented atlases lacking semantic coherence, significantly limiting their utility for tasks requiring meaningful part decomposition (e.g., texture map editing).
Nuvo~\cite{srinivasan2024nuvo} and FAM~\cite{zhang2024flattenanything} attempt to improve the continuity of atlases through neural field-based parameterization. 
However, these methods still struggle to produce globally coherent cuts that align with semantic or functional boundaries.
Recent learning-based surface segmentation methods, such as PartField~\cite{zhu2023nerve}, can provide semantic guidance for part decomposition but often fail to produce clean boundaries for cutting.
Alternatively, existing wire-frame detection methods~\cite{gori2017flowrep, zhu2023nerve, CLR-Wire} are limited to simple scenarios (e.g., CAD models). 
% and often fail to produce globally continuous and topologically consistent cuts.
% key challenge

\textit{Where to put the cuts?}
The optimal cuts should 1) minimize geometric distortion for UV parameterization, 2) preserve semantically meaningful boundaries, and 3) maintain functional continuity for downstream applications.
In professional practice, artists often manually define mesh cuts in interactive workflows: they init cutting at some topological singularities, and gradually connect more points to form cutting seams in a sequential way, and prioritize functionally significant boundaries.
The recent release of large-scale 3D artistic mesh datasets (e.g., Objaverse~\cite{objaverse} and Objaverse-XL~\cite{objaverseXL}) provides extensive artist-created cutting seam annotations.
This motivates us to learn the distribution of artist-crafted seams to automate semantically meaningful surface cutting.

Auto-regressive modeling has achieved remarkable success in content generation across different modalities, such as Large Language Models (LLMs)~\cite{achiam2023gpt}, image generation~\cite{VAR}, and polygon mesh generation~\cite{hao2024meshtron}.
We find that auto-regressive modeling is well-suited to the surface cutting task. It mimics the casual decision-making process of manual cutting by capturing the sequential dependencies between successive cuts.

% our approach

To this end, we introduce SeamGPT, an auto-regressive model that generates artist-style cutting seams, as sequences of 3D line segments.
Specifically, given an input surface mesh, we sample point clouds on vertices and edges and compress them into a latent shape condition using a point cloud encoder. 
A GPT-style transformer decoder then auto-regressively generates cutting seams in the form of line segments.
Each seam segment is represented by a start and an end 3D point with quantized coordinates, produced in axis-sorted order.
We train SeamGPT on a curated dataset of 560K artist-cut meshes, filtered to retain only those with artist-annotated seams exhibiting clear semantic intent.
Experiments across multiple mesh categories demonstrate that our method significantly improves surface cutting quality for UV-unwrapping.
In addition, in certain scenarios, the generated seams can serve as effective boundary indicators for 3D part segmentation, providing a potential solution for advancing 3D segmentation.

Our key contributions are as follows:
\begin{itemize}
\item We present a new auto-regressive formulation for cutting seams as a sequence of line segments, tailoring a GPT-inspired  transformer, to produce semantically meaningful and consistent artist-style cutting.
\item For UV-unwrapping, SeamGPT achieves state-of-the-art performance on benchmarks containing both manifold and non-manifold meshes, including artist-created meshes and AI-generated meshes (e.g., from latent 3D diffusion models and polygon-generation methods).  
\item For part segmentation, SeamGPT complements existing tools (e.g., Parfield~\cite{partfield2025}) by providing clean boundaries, addressing a key limitation in current boundary refinement pipelines.  
\end{itemize}

\section{Related Work}

\textbf{Mesh Cutting for Flattening.}
Mesh cutting represents a well-established research problem in computer graphics.
The primary use of mesh cutting is to achieve low-distortion mesh flattening for texture mapping.
Cutting strategies are broadly categorized as bottom-up or top-down.
Bottom-up approaches first partition the surface into small, compact regions and iteratively merge connected regions~\cite{sorkine2002bounded, zhou2004iso-charts, yamauchi2005mesh}.
Sorkine et al.~\cite{sorkine2002bounded} pioneered this paradigm by expanding seed triangles until reaching distortion bounds, later refined in~\cite{zhou2004iso-charts, yamauchi2005mesh}.
These methods form the foundation for industrial tools like xAtlas and Blender’s Smart UV Unwrapper.
While technically valid, they often produce over-fragmented atlases lacking semantic coherence, which limits their utility for texture editing workflows.
Top-down methods cut the surface with seams or winding boundaries, either through user-guided or automatic techniques~\cite{mitani2004making, tang2016interactive, sheffer2002seamster}.
Cutting seams can be automatically generated by connecting cone singularities~\cite{kharevych2006discrete, springborn2008conformal, soliman2018optimal-con-singularities}.
After defining seams, the flattening process typically relies on distortion optimization techniques based on mesh connectivity.
Least Squares Conformal Mapping (LSCM)~\cite{levy2023least} is widely used for generating conformal maps for pre-cut meshes. 
Variational Surface Cutting~\cite{Sharp:2018:VSC} presents a global variational approach for low-distortion cuts, using conformal shape derivative flow for parameterization-free flattening with user constraints. %
Geometry Images~\cite{gu2002geometry} and Multi-Chart Geometry Images~\cite{sander2003multi-chart} encode 3D meshes as regular grids for efficient storage and re-meshing.
Rectangular Multi-Chart Geometry Images~\cite{carr2006rectangular} enabled GPU-friendly regular grids.
OptCuts~\cite{li2018optcuts} jointly optimize cut length and distortion through alternating mapping and cut locus optimization; the process involves hybridized discrete/continuous optimization.
SeamCut~\cite{Lucquin:2017:SeamCut} automates segmentation via field-based analysis while remaining topology-agnostic.
Recent neural field-based approaches leverage continuous MLP functions to model cutting and UV mapping.
AtlasNet~\cite{groueix2018atlasnet} pioneered this direction by using neural fields to represent a continuous mapping from 2D atlases to 3D surfaces.
Nuvo~\cite{srinivasan2024nuvo} optimizes multiple neural fields for UV unwrapping with explicit parameterization constraints.
FAM~\cite{zhang2024flattenanything} extend this direction with interpretable sub-networks surface cutting, UV deforming, unwrapping, and wrapping, the sub networks are assembled into a bi-directional cycle mapping framework.
While these methods generate continuous mappings, they typically require per-scene optimization and often disregard object semantics.

\textbf{3D Part Segmentation.}
Traditional data-driven part segmentation approaches require predefined part templates, follow a supervised learning pipeline, and suffer from the limited scale and diversity of 3D part-annotated datasets.
Due to the success of 2D foundation models like SAM~\cite{kirillov2023sam}, recent open-world 3D segmentation has made significant progress.
Various methods have been developed to lift and merge multi-view 2D SAM predictions for both scene-level and object-level 3D segmentation. 
For instance, Part123~\cite{liu2024part123} integrates SAM into 3D reconstruction, enabling part-aware single-image object generation.
Similarly, SAMPart3D~\cite{yang2024sampart3d} employs per-shape optimization to distill 2D segmentations into 3D.
However, these approaches rely on multi-step inference pipelines with lengthy optimizations. 
In contrast, PartField~\cite{partfield2025} proposes a single-stage feedforward method for learning part-based 3D features.
HOLOPart~\cite{yang2025holopart} further refines segmented parts using a 3D diffusion model that captures both part geometry and global shape information—though the completion quality heavily depends on the initial segmentation.
A key limitation of existing 3D part segmentation methods is their inability to produce clean part boundaries. 
In this paper, we demonstrate that our auto-regressive surface-cutting approach has the  potential to alleviates this issue by providing precise cutting seams as clean part boundaries.

\textbf{Line and wireframe reconstrution.}
Reconstructing lines or wireframes from input point clouds or meshes is a closely related task.
Traditional line feature detection in point clouds primarily relies on local geometric features, such as the eigen-structure of the covariance matrix, normals, curvatures, or statistical metrics~\cite{bazazian2015fast,daniels2007robust,lin2015line,hackel2016contour}.
With the advent of deep learning, several works~\cite{wang2020pienet,bazazian2021edc,himeur2021pcednet} leverage neural networks to  detect edge point and formulate it as a per-point classification task.
EC-Net~\cite{yu2018ecnet} reformulates this as a regression problem, learning residual point coordinates and point-to-edge distances to identify edge points.
PointNet~\cite{qi2017pointnet} and 3D graph convolutions~\cite{wang2019dynamic} are widely adopted as baselines for point feature detection.
The next step extracting parametric curves from detected points or lines.
As done in~\cite{gumhold2001feature,wang2020pienet}, a common strategy is grouping and connecting the detected points, then determining the target parametric curve type for fitting.
NerVE~\cite{zhu2023nerve} leverages piece-wise linear (PWL) curve representations to enable efficient parametric curve extraction.

\section{Data Preparation}
\textbf{Data collection and filtering.} 
We collected our training data from several open-source 3D datasets, including Objaverse~\cite{objaverse}, Objaverse-XL~\cite{objaverseXL}, and 3D-FUTURE~\cite{fu20213d}, focusing primarily on meshes with existing UV coordinates that contain essential artist-defined seam information. To ensure high-quality training data, we implemented a rigorous filtering process that excluded meshes with poor topology such as raw 3D scans and low-quality AI-generated models. We removed problematic UV unwrappings such as those with overlapping UVs or arbitrarily placed seams lacking semantic meaning. Through this comprehensive filtering process, we assembled a training dataset of 1.9 million objects for initial training, then further refined our approach by fine-tuning on a high-quality subset of approximately 560K meshes featuring exemplary artist-created seams. 

\textbf{Mesh Seam Extraction.} To extract ground-truth cutting seams from our filtered meshes, we leveraged the existing UV parameterization information. The boundaries between UV islands naturally define the cutting seams on the original mesh. We identified all UV island boundary edges and projected them back onto the 3D mesh to obtain the corresponding cutting seams. These edges were then converted into our sequential representation format as ordered sequences of line segments.

\section{Method}  
\label{ sec:seamgpt}

\begin{figure*}[!t]
    \centering
    \includegraphics[width=\textwidth]{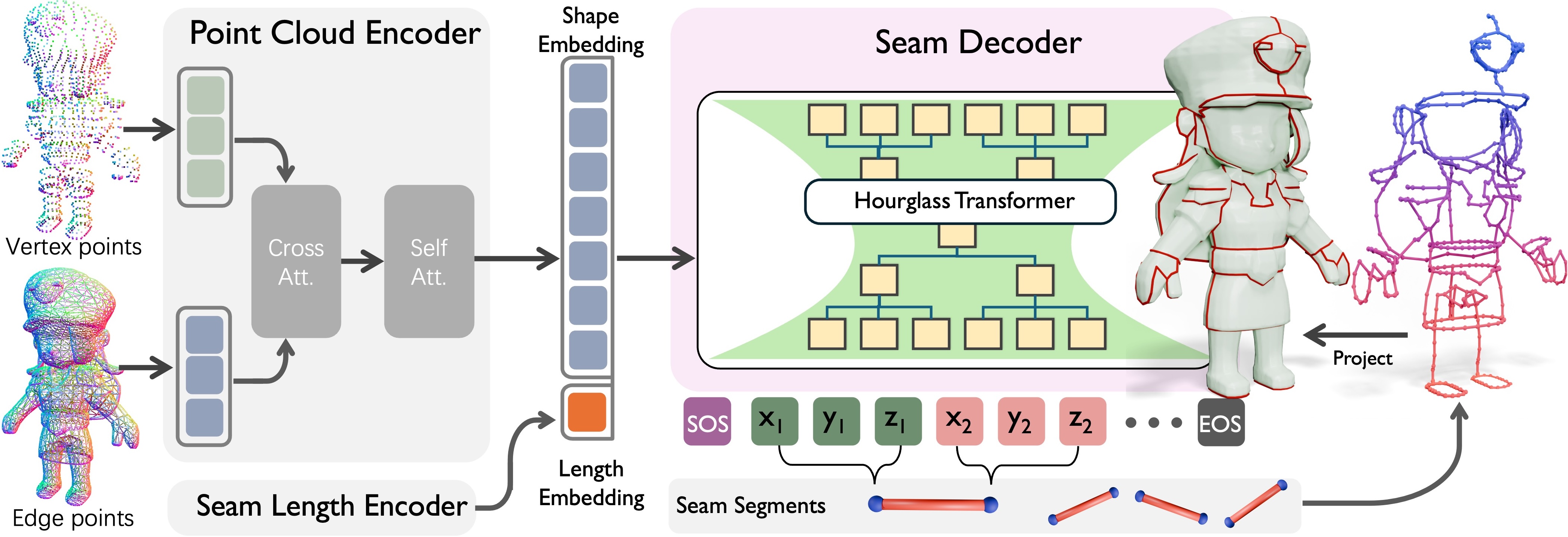}
    \caption{SeamGPT architecture:  Point cloud encoder extracts shape context; Causal transformer decoder generates axis-ordered seam coordinates. 
    Color indicates the prediction order is of the seam segments (red to blue).
    }
    \label{fig:pipeline}
\end{figure*}

We introduce SeamGPT, a novel framework that generates artist-style cutting seams through an auto-regressive approach. Our method formulates surface cutting as a sequence prediction problem, where cutting seams are represented as an ordered series of 3D line segments. Given an input mesh $\mathit{M}$, our goal is to generate seam edges $S = \{s^{i}\}_{i \in [N_s]}$. The overview of SeamGPT is shown in Fig.~\ref{fig:pipeline}. We first introduce our seam representation strategy in Sec.~\ref{sec.4.1}, which encodes cutting seams as sequential tokens. In Sec.~\ref{sec.4.2}, we detail our auto-regressive generation process, which mimics the sequential decision-making of professional artists. 
% Sec.~\ref{sec.4.3} presents our training strategy.

\subsection{Mesh Seam Representation}
\label{sec.4.1}
A seam sequence $S$ of $N_s$ segments $\{s^i\}_{i \in [N_s]}$ is defined as: $S = \{s^1, s^2, \ldots s^{N_s}\}$, where each segment $s^i$ is a 3D line segment represented by two vertices: $s^i = (p^i_h, p^i_t)$, i.e. head and tail. Each vertex $p$ is defined by its 3D coordinates: $p = (x, y, z)$. 
Thus, a seam sequence can be decomposed at multiple levels:
\begin{align}
 S &= \{s^1, s^2, \ldots s^{N_s}\} && \mathrm{Segment\quad level} \nonumber\\
 &= \{p^1_h, p^1_t, p^2_h, p^2_t, \ldots, p^{N_h}_t, p^{N_h}_t\} && \mathrm{Point \quad level} \label{eq:seam_levels} \\
 &= \{x^1_h, y^1_h, z^1_h, x^1_t, y^1_t, z^1_t, \ldots, x^{N_s}_t, y^{N_s}_t, z^{N_s}_t\} && \mathrm{Coord. \quad level} \nonumber
\end{align}
\textbf{Seam ordering.}
For an auto-regressive model to function properly, a consistent order of sequences is required. 
Following existing practice for mesh generation~\cite{siddiqui2023meshgpt, bpt, hao2024meshtron},
we first sort vertices $yzx$ order, where $y$ represents the vertical axis, and then sort two vertices within an edge lexicographically, placing the lowest $yzx$-ordered vertex first. 
Finally, seam edges are sorted in ascending $yzx$-order based on the sorted values of their vertices.
The resulting order can be seen through the color coding of the generated meshes presented in Figure~\ref{fig:pipeline}, i.e. from red to blue.

\textbf{Quantization of coordinates.} Autoregressive models typically sample from a multinomial distribution over a discrete set of possible values. To adhere to this convention, we quantize vertex coordinates into a fixed number of discrete bins. The quantization resolution—determined by the number of bins—directly affects the precision of the predicted seam. Higher quantization levels yield more detailed and accurate representations but also increase the complexity of the generation process. To balance precision and tractability, we employ 1024-level quantization, enabling effective representation of complex seams.

\subsection{Autoregressive Seam Prediction}
\label{sec.4.2}

In autoregressive seam prediction, a seam sequence $S$ is generated by sequentially predicting each coordinate $c_i$ based on its conditional probability given all previously generated coordinates $P(c_i | c_{<i})$. The probability of the entire seam is then given by the joint probability of all its coordinates:
\begin{equation}
    P(S) = \prod_{i=1}^{6N_s} P(c_i | c_{<i}).
    \label{eq:seam_arm}
\end{equation}
\textbf{Global Shape Conditioning.}
Point clouds are a flexible and universal 3D representation that can be efficiently derived from other 3D formats, including meshes.
We use a point cloud encoder to extract representative features for characterizing the input 3D shapes.
In the context of surface cutting, seams are encouraged to align with the vertices and edges of the original mesh, such that cutting the mesh along seams does not create excessive extra faces.
To guide the decoder in producing vertex and edge-aligned seam placement, instead of sampling point clouds uniformly, we sample surface points only on vertices and along edges.
Specifically, we sample a total of 61,440 points, evenly split between: 30,720 points on vertices and 30,720 points on edges.
If the input mesh has fewer than 30,720 vertices, we use repeated over-sampling.
Points along an edge are sampled uniformly by interpolating between its start and end points with K samples, where K is determined based on the edge's length.
Finally, the input points are fed into a jointly trained point cloud encoder from~\cite{hunyuan3d22025tencent}, which processes the point cloud through a series of cross- and self-attention layers and compresses the point cloud to a latent shape embedding of length 3072 and dimension 1024.
Another option to create shape embeddings is to use mesh encoders, such as~\cite{zhou2020fullymeshae}. However, the computational cost of mesh encoder does not scale well when the input has a large number of vertices. We show in the ablation study that point cloud conditioning produces much better results than mesh conditioning.

\textbf{Seam Count control.}
Given an input shape, multiple possible suitable cutting solutions exist.
Depending on the application requirements, one can make many cuts to decompose a mesh, or just a few.
To regulate the cutting granularity, we concatenate a length embedding to the shape embedding.
We find that modulating the length embedding directly controls cutting granularity.

\textbf{HourGlass Decoder Architecture.}
Following~\cite{hao2024meshtron}, we build an hourglass-like autoregressive decoder architecture to sequence at multiple levels of abstraction.
The architecture employs multiple Transformer stacks at each level, with transitions between levels managed by causality-preserving shortening and upsampling layers that bridge these hierarchical stages.
There are three hierarchical levels: coordinates, vertices, and edges.
The input coordinate sequence is shortened by a factor of 3 at the vertex level.
It is then further shortened by a factor of 2 at the edge level.
Both shortening and upsampling layers are implemented to preserve causality.
The expanded sequence is combined with higher-resolution sequences from earlier levels via residual connections, similar to U-Nets.

\textbf{Training Strategy.} We employ two loss functions to for model training: a cross-entropy loss for token prediction and a KL-divergence loss to regularize the shape embedding space, ensuring it remains compact and continuous. 
Training begins with a 2,000-step warm-up phase and is parallelized across 64 Nvidia H20 GPUs (98GB Mem.) with a total batch size of 128. 
The model converges after one week of training. During training, we first scale all samples to fit within a cubic bounding box in the range of $-1$ to $1$.
We then apply data augmentation techniques, including random scaling within $[0.95,1.05]$, random vertex jitter, and random rotation.
\section{Mesh UV-unwrapping Experiment}

\begin{figure}[!t]
    \centering
    \includegraphics[width=1\textwidth]{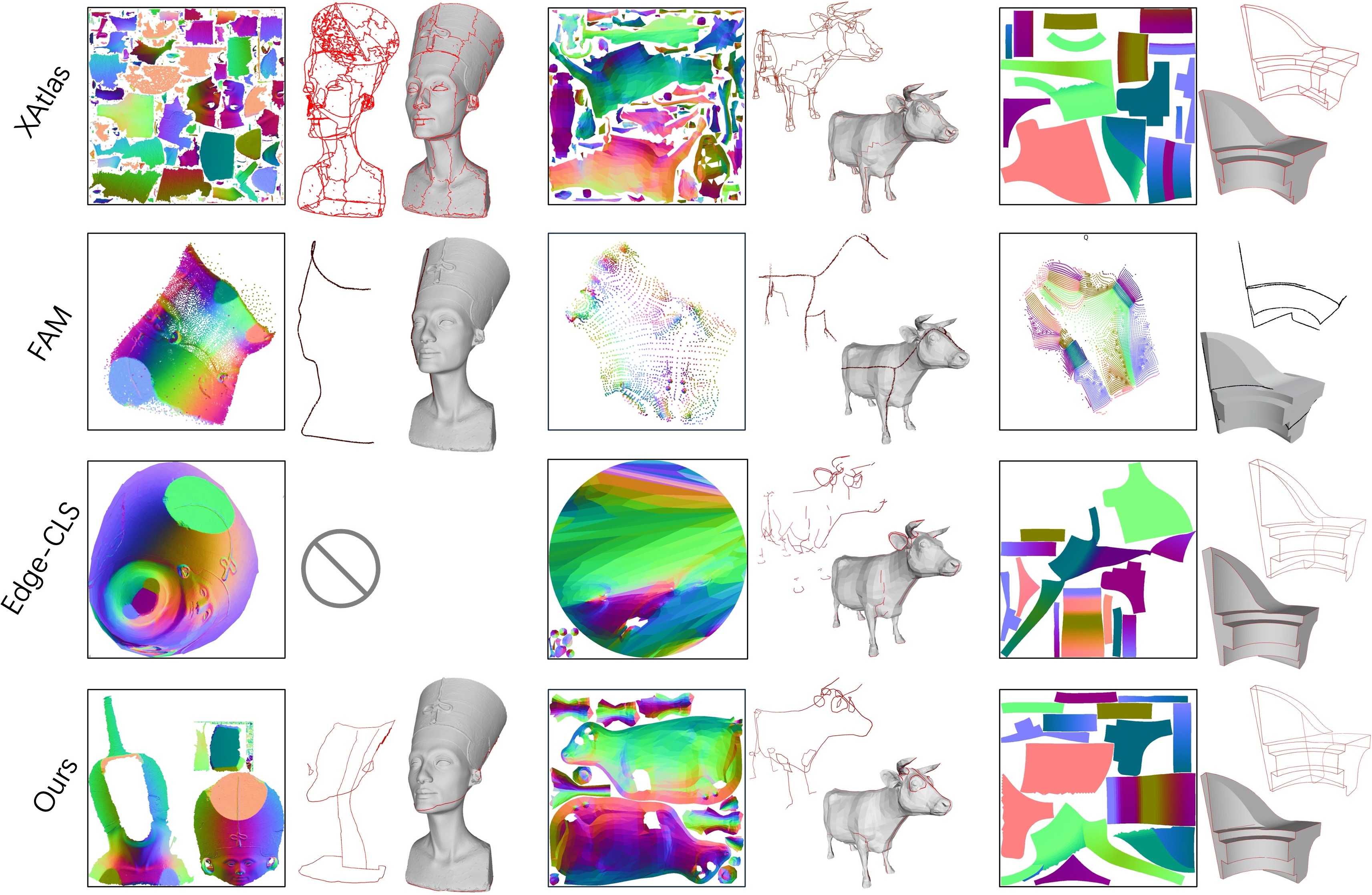}
\caption{Qualitative UV flatten results on FAM benchmark (  Nefertiti, Cow, and Fandisk). }
\label{fig:fam-compare}
\end{figure}

\textbf{ Benchmarks and Evaluation Metric.}
We conduct experiments on a diverse collection of 3D surface models from Flatten Anything (FAM)~\cite{zhang2024flattenanything}, which primarily includes low-poly meshes, CAD models, and 3D scanned meshes. We also evaluate on Toys4K~\cite{stojanov21cvpr}, a dataset of non-manifold artist-created meshes.  
Our evaluation leverages the \textbf{Mesh distortion} metrics, which is computed as the average conformal energy over all triangular faces of the mesh.

\textbf{Baselines and Implementations.} We compare SeamGPT against several state-of-the-art methods for mesh UV-unwrapping. \textbf{XAtlas}~\cite{xatlas} employs a bottom-up approach with bounded distortion charts. \textbf{Nuvo}~\cite{srinivasan2024nuvo} leverages neural fields with explicit parameterization constraints. \textbf{FAM}~\cite{zhang2024flattenanything} implements interpretable sub-networks in a bi-directional cycle mapping framework. 
We also built another baseline called \textbf{Edge-CLS}, which takes a mesh as input and uses graph convolution and Transformer layers to compute per-edge features. These features are then fed into an MLP classifier to predict whether each edge is a seam edge or not (i.e., this is an edge classification baseline).
We train Edge-CLS on the same training set and use the same UV-unwrapping process as SeamGPT.

\textbf{SeamGPT-based UV-unwrapping.} Once SeamGPT generates cutting seams, we implement a streamlined unwrapping process to create practical UV maps. We first map each predicted seam point to its nearest vertex on the input mesh, then connect these vertices through shortest geodesic paths along the mesh edges. We then cut the mesh by duplicating vertices along these paths, creating independent boundaries for flattening. Finally, we apply Blender's Minimum Stretch algorithm to the segmented mesh, optimizing UV coordinates to evenly distribute stretching while preserving the semantic structure defined by our seams. This process yields low-distortion UV mappings that respect functional and aesthetic boundaries, improving upon conventional automated methods.

\textbf{Comparison results.}
Results. Tables~\ref{fam-bench} and~\ref{toys4k-bench}, along with Figure~\ref{fig:fam-compare}, present qualitative and quantitative results.
SeamGPT achieves the best performance across all metrics. In contrast: XAtlas generates over-fragmented cuts, FAM fails to produce subtle cuts consistently, Edge-CLS performs well only on sharp edge features but struggles with generating seams on smooth, featureless regions.
Our method consistently produces semantic and reasonable cuts regardless of surface characteristics.

\begin{table}[!ht]
    \centering
 
    \resizebox{\textwidth}{!}{%  % Scales width to \textwidth, height proportionally
    \renewcommand{\arraystretch}{1.5}
    \begin{tabular}{lccccccccccccc|c}
    \toprule
        
         ~ & Bimba & Lucy & Ogre & Armadi. & Bunny & Nefert. & Dragon & Homer & Happy & Fandi. & Spot & Arm & Cow & Avg. \\ \hline
        Xatalas~\cite{xatlas} & 15.44  & \textbf{0.01 } & \textbf{0.66}  & \textbf{0.17}  & 61.84  & \textbf{0.03}  & \textbf{0.22}  & \textbf{7.51}  & 99.84  & 8.41  & 12.77  & 29.98  & \textbf{1.94}  & 18.37  \\ \hline
        Nuvo~\cite{srinivasan2024nuvo} & 19.12  & 57.89  & 26.22  & 114.21  & 16.84  & 20.92  & 61.03  & 21.92  & 267.43  & 19.76  & 12.93  & 37.34  & 12.70  & 52.95  \\ \hline
        FAM~\cite{zhang2024flattenanything} & 12.10  & 35.14  & 11.55  & 59.87  & 7.33  & 11.21  & 904.89  & 14.19  & 23.00  & 12.21  & 9.37  & 20.98  & 8.49  & 86.95  \\ \hline
        Edge-CLS  & \textbf{8.14}   & 22.85   & 23.54   & 11.18   & \textbf{3.91}   & 4.67   & 15.39   & 20.16   & \textbf{27.37}   & 12.47   & \textbf{5.77}   & 87.20   & 9.20   & 19.37  \\ \hline
        Ours & 10.68  & \textbf{0.01}  & 2.01  & 2.47  & 50.47  & 0.12  & 0.56  & 10.28  & 61.68  & \textbf{8.15}  & 5.95  & \textbf{14.88}  & 2.24  & \textbf{13.04}  \\ \bottomrule

    \end{tabular}
    }
    \caption{Quantitative results on Flatten-Anything benchmark using the face distortion metric.}
    \label{fam-bench}

\end{table}

% \begin{table}[!ht]
%     \centering
%     \begin{tabular}{lllllllllllllll}
%         ~ & Bimba & Lucy & Ogre & Armadi. & Bunny & Nefert. & Dragon & Homer & Happy & Chebu. & Spot & Arm & Cow & Avg. \\ 
%         Xatalas & 15.44  & 0.01  & 0.66  & 0.17  & 61.84  & 0.03  & 0.22  & 7.51  & 99.84  & 8.41  & 12.77  & 29.98  & 1.94  & 18.37  \\ 
%         Nuvo\~$\backslash$cite\{srinivasan2024nuvo} & 19.12  & 57.89  & 26.22  & 114.21  & 16.84  & 20.92  & 61.03  & 21.92  & 267.43  & 19.76  & 12.93  & 37.34  & 12.70  & 52.95  \\ 
%         FAM\~$\backslash$cite\{zhang2024flattenanything} & 12.10  & 35.14  & 11.55  & 59.87  & 7.33  & 11.21  & 904.89  & 14.19  & 23.00  & 12.21  & 9.37  & 20.98  & 8.49  & 86.95  \\ 
%         Ours & 10.68  & 0.01  & 2.01  & 2.47  & 50.47  & 0.12  & 0.56  & 10.28  & 61.68  & 8.15  & 5.95  & 14.88  & 2.24  & 13.04  \\ 
%         Edge-CLS  & 8.14   & 22.85   & 23.54   & 11.18   & 3.91   & 4.67   & 15.39   & 20.16   & 27.37   & 12.47   & 5.77   & 87.20   & 9.20   & 19.37  \\ 
%     \end{tabular}
% \end{table}

\begin{table}[!ht]
    \centering

    \resizebox{0.8\textwidth}{!}{%  % Scales width to \textwidth, height proportionally
    \renewcommand{\arraystretch}{1.35}
    \begin{tabular}{lcccccccc|c}
    \toprule
    
&Bowl  & Ball  & Sheep  & Driver  & Chicken  & Apple  & Giraffe  & Bottle  & Avg.   \\ \hline

Xatalas~\cite{xatlas} &       0.91  &\textbf{ 0.26}  & \textbf{1.19}  & 4.61  & 2.36  & \textbf{3.11}  & 2.85  &\textbf{ 0.57}  & 1.98   \\ \hline
Nuvo~\cite{srinivasan2024nuvo}  &      3.99  & 1.33  & 10.43  & 33.07  & 9.79  & 15.39  & 21.04  & 6.02  & 12.63    \\ \hline
FAM~\cite{zhang2024flattenanything}   &     3.80  & 0.81  & 6.45  & 15.33  & 18.98  & 6.64  & 11.77  & 4.36  & 8.52    \\ \hline
Ours &    \textbf{0.49 }  & 0.31   & 1.39   &\textbf{ 4.25}   &\textbf{ 1.86}   & 4.02   & \textbf{2.59}   & 0.67   &\textbf{ 1.95}    \\ \bottomrule

    \end{tabular}
    }
    \caption{Quantitative results on Toys4K Benchmark using the face distortion metric.}
    \label{toys4k-bench}

\end{table}

\textbf{User study.} To further assess our method's practical utility, we conducted a user study with 20 professional 3D artists evaluating \textbf{Boundary} quality and \textbf{Editability}. Boundary quality measures how unfragmented a UV map is, while editability reflects how well the mapping supports appearance editing. Participants rated UV unwrappings from all methods on a 5-point scale. As shown in Table ~\ref{tab:userstudy}, SeamGPT significantly outperforms existing methods in both metrics.

\begin{table}[!ht]

\centering
\resizebox{\textwidth}{!}
{
\small
\renewcommand{\arraystretch}{1.2}

\begin{tabular}{l|cc|cc|cc|cc}
\toprule
\multicolumn{1}{c}{} & \multicolumn{2}{|c}{\texttt{Fandisk}} & \multicolumn{2}{|c}{ Cow} & \multicolumn{2}{|c}{\texttt{Nefertiti}} & \multicolumn{2}{|c}{ Avg. }\\
\multicolumn{1}{c|}{} & 
\multicolumn{1}{c}{Boundary $\uparrow$} & \multicolumn{1}{c|}{Editability $\uparrow$} & \multicolumn{1}{c}{Boundary $\uparrow$} & \multicolumn{1}{c|}{Editability $\uparrow$} & \multicolumn{1}{c}{Boundary $\uparrow$} & \multicolumn{1}{c|}{Editability $\uparrow$} & \multicolumn{1}{c}{Boundary $\uparrow$} & \multicolumn{1}{c}{Editability $\uparrow$} \\ \hline

Xatalas~\cite{xatlas}
& \textbf{4.42} & \textbf{4.42} & 2.68  & 2.37 & 2.79  & 2.47 & 3.30 & 3.09 \\ \hline

Nuvo~\cite{srinivasan2024nuvo}
& 1.32  & 1.32  & 1.16  & 1.21  & 1.42 & 1.42 & 1.30 & 1.32 \\ \hline

FAM~\cite{zhang2024flattenanything}
& 1.74  & 1.53 & 1.84  & 1.53 & 2.05  & 1.84 & 1.88 & 1.63 \\ \hline

Edge-CLS
& 3.89  & 3.84 & 2.79  & 2.32 & 2.58 & 2.16 & 3.09 & 2.77 \\ \hline

Ours
& 4.37  & 4.32 & \textbf{4.16}  & \textbf{4.16} & \textbf{3.47}  & \textbf{3.58} & \textbf{4.00} & \textbf{4.02} \\ 
\bottomrule
\end{tabular}
}
\caption{User Study about Boundary quality and Editability.}
\label{tab:userstudy}
\end{table}

\section{Ablation Study}

\begin{wrapfigure}{r} {0.5\textwidth}
\centering
\begin{center}
\includegraphics[width=\linewidth]{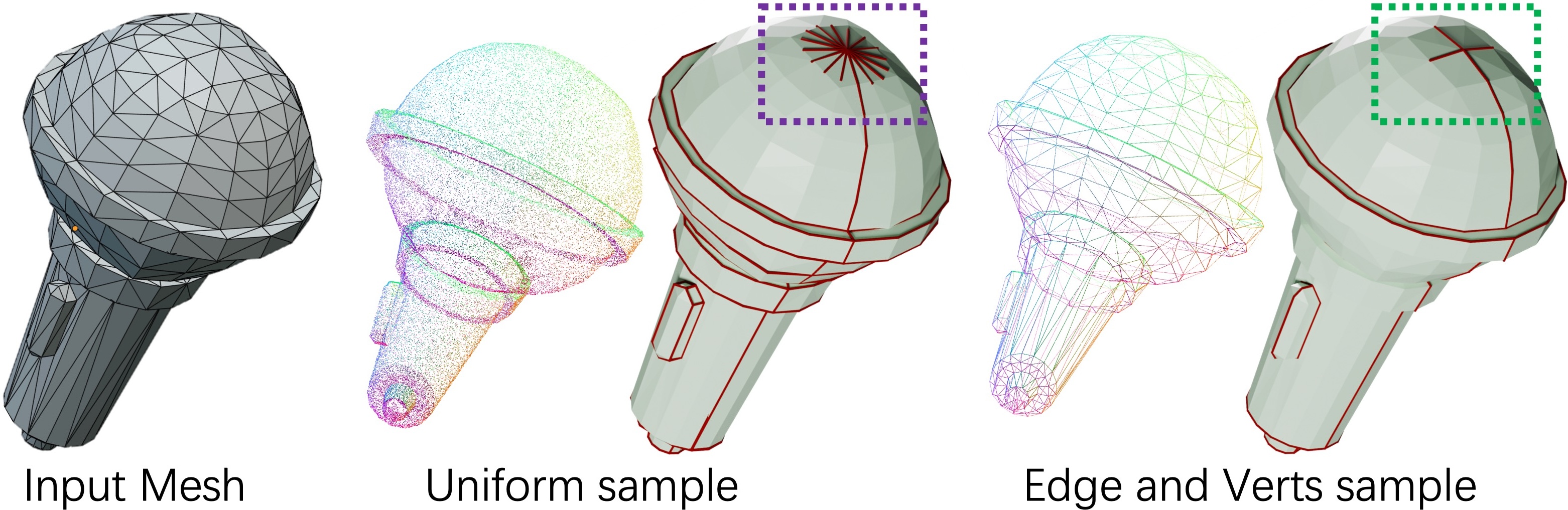}
\end{center}
\caption{Ablation of point sampling strategy.}
\label{fig:ablation-point-sampling}
\end{wrapfigure}

\textbf{Point cloud sampling strategy.} As shown in Figure.~\ref{fig:ablation-point-sampling}, when conditioned on point clouds uniformly sampled across the mesh surface, the generated seams remain logically valid from a surface-cutting perspective but may not precisely align with the input mesh's vertices and edges. 
In contrast, sampling point clouds along edges and vertices produces seams that naturally conform to the mesh topologies. This could prevents creating excessive extra mesh faces.
We also found that sampling along edges and vertices significantly improves model convergence, as the transformer gains explicit positional awareness of potential cutting coordinates.

\begin{wrapfigure}{r}[0.05\textwidth]{0.5\textwidth}  % r 表示右对齐，0.5\textwidth 是图片宽度
\centering
\begin{center}
\includegraphics[width=\linewidth]{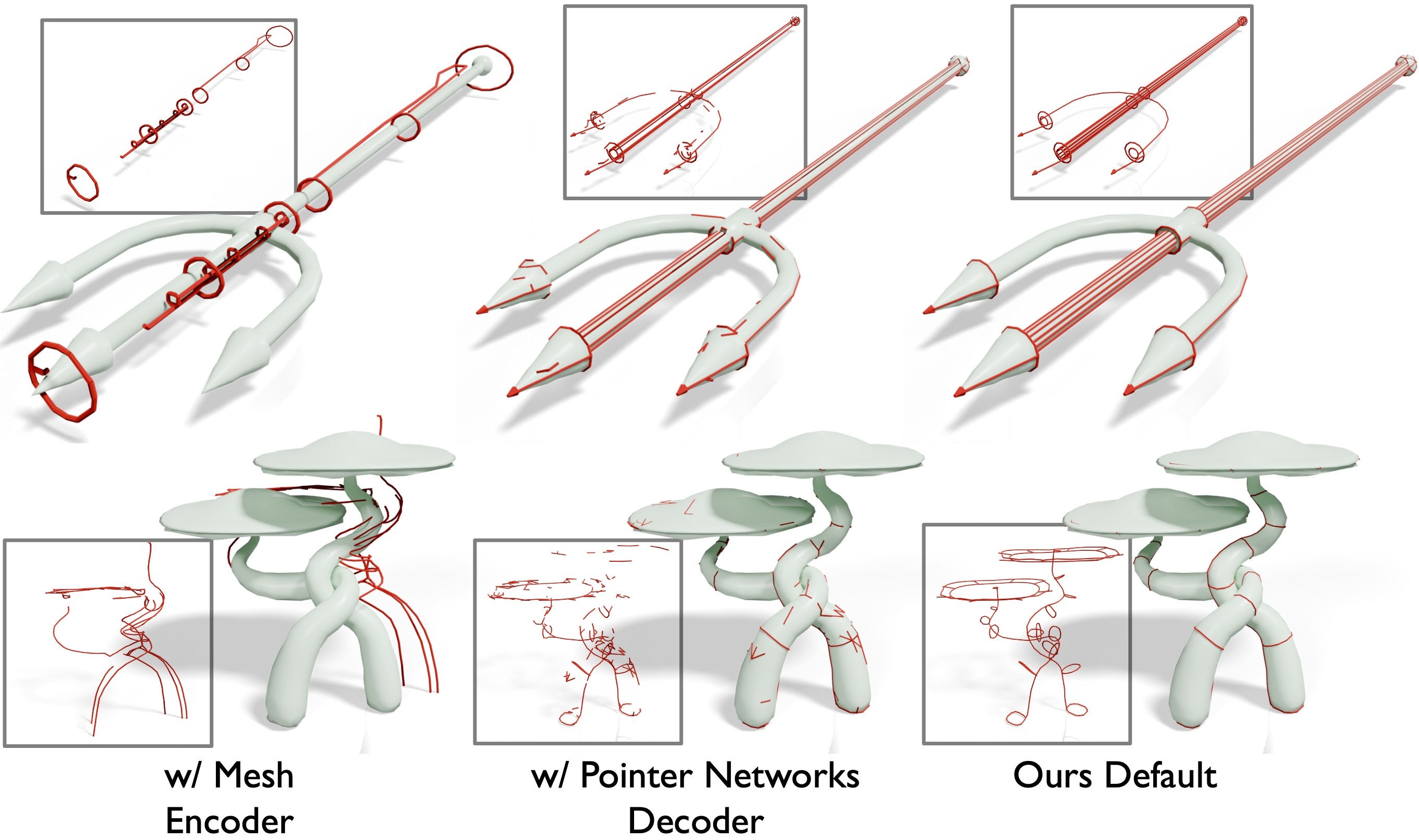}
\end{center}
\begin{minipage}{0.5\textwidth}  % 标题宽度限制为0.5\textwidth
% \raggedleft
    \caption{Ablation study of encoder and decoder.}
    \label{fig:ablation-ptrnet-meshcond}
\end{minipage}
\end{wrapfigure}

\textbf{Mesh encoder vs Point cloud encoder.}  An alternative approach for generating shape embeddings employs mesh encoders, as demonstrated by Zhou et al.~\cite{zhou2020fullymeshae}.
We implemented an encoder combining graph convolutions (operating on both vertices and edges) with a full self-attention transformer. This encoder produces vertex-wise tokens that are subsequently fed to the decoder via cross-attention mechanisms.
As shown in Figure~\ref{fig:ablation-ptrnet-meshcond}, point-cloud encoder yields superior results compared to mesh encoders. Furthermore, the computational cost of our mesh encoder scales poorly with increasing vertex counts.
Mesh encoder-based methods often fail to accurately capture the precise positions of original vertices, resulting in significant misalignment between the generated seam edges and the original mesh.

\textbf{Does Pointer networks works?}
In the case that the cutting seam forms a subset of the edges in the mesh, we can also adopt the Pointer Network~\cite{vinyals2015pointer-networks} architecture, which auto-regressively produce the pointers to the mesh edges.
We follow the implementation of Polygen~\cite{nash2020polygen} to build a pointer network with a mesh encoder that produces edge-wise embedding and a casual transformer to create pointers to the edges that lie on the seams auto-regressively. 
Pointer network struggles to generate consistent seams, often resulting in discontinuous cuts as demonstrated in Figure~\ref{fig:ablation-ptrnet-meshcond}.

\begin{figure}[!h]
    \centering
    \includegraphics[width=1\textwidth]{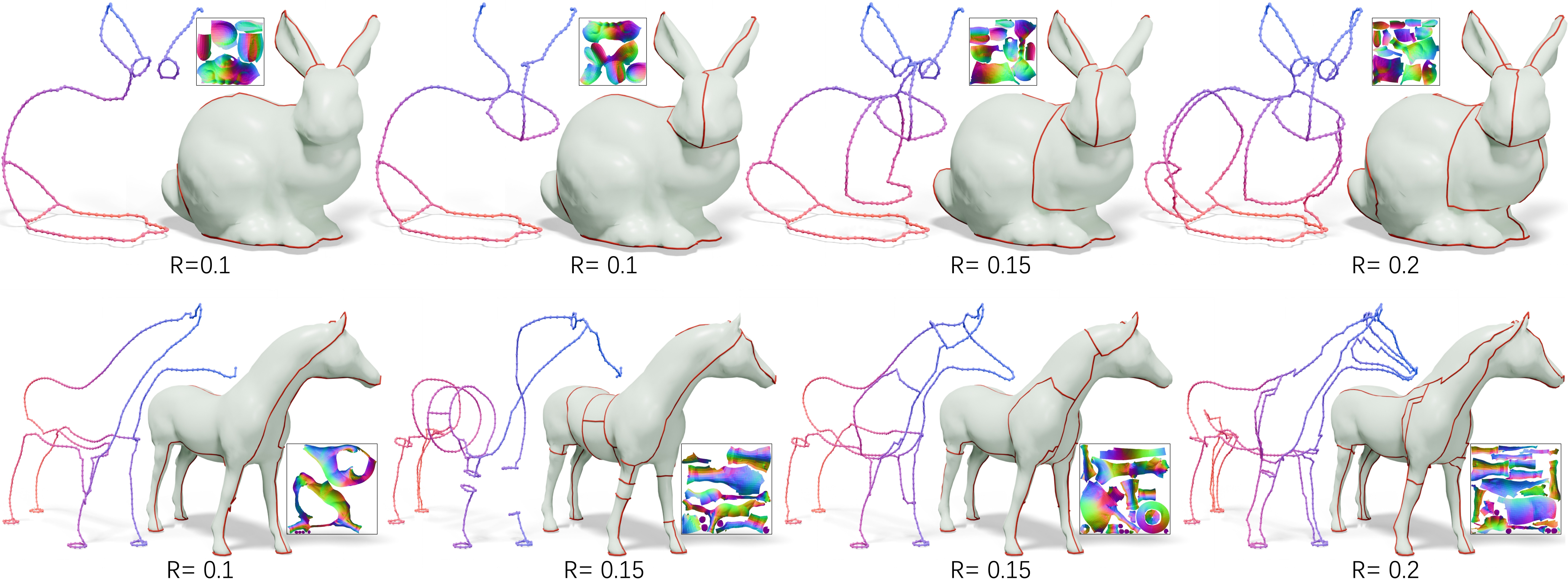}
\caption{Seam length control and diversity. We can control the cutting granularity by adjusting seam length. Diverse valid cutting seams can be generated.}
\label{fig:bunny-alation}
\end{figure}

\textbf{Seam length control and diversity.}  We define R as the ratio of seam segment count to the number of mesh vertices. Empirically, valid cutting seams typically has R value within the range [0.1,0.35]. Above this range result in over-cutting, while values below it lead to insufficient cuts. As shown in Figure~\ref{fig:bunny-alation}, controlling R allows us to adjust the granularity of the cuts.
Additionally, due to the non-deterministic nature of autoregressive transformers, we can generate diverse valid cutting seams from the same length control.

\section{Seam Enhanced 3D Part Segmentation}
\label{sec.5.2}
In this section, we explore the potential of our SeamGPT to enhance part segmentation. 3D part segmentation is a long-standing problem in 3D computer vision, aiming to separate a 3D object into meaningful components. With the success of large 2D foundation models such as SAM~\cite{kirillov2023sam}, recent works have explored leveraging 2D priors to enable open-world 3D segmentation capabilities. For example, SAMPart3D~\cite{yang2024sampart3d} employs 3D pretraining to extract multi-view features into a 3D encoder and applies per-shape optimization for part segmentation based on multi-view SAM predictions. PartField~\cite{partfield2025} trains a feedforward model supervised by multi-view masks to predict a part-based feature field for 3D shapes. The learned features can then be clustered to facilitate part decomposition. However, both SAMPart3D and PartField leverage contrastive learning to distill multi-view features for part segmentation. One of the primary problems with multi-view distillation is the potential for blurriness along the edges of segmented parts, which can lead to ambiguous transitions and make it difficult to obtain clear and well-defined components. This issue often arises from the varying perspectives inherent in multi-view renderings and the limitations of the model in accurately delineating boundaries between distinct segments. To address this, we propose utilizing the seam lines generated by SeamGPT to complement existing tools (e.g., PartField) and provide clean boundaries for part segmentation.

\begin{figure}[!h]
    \centering
    \label{fig:segvis}
    \includegraphics[width=1\textwidth]{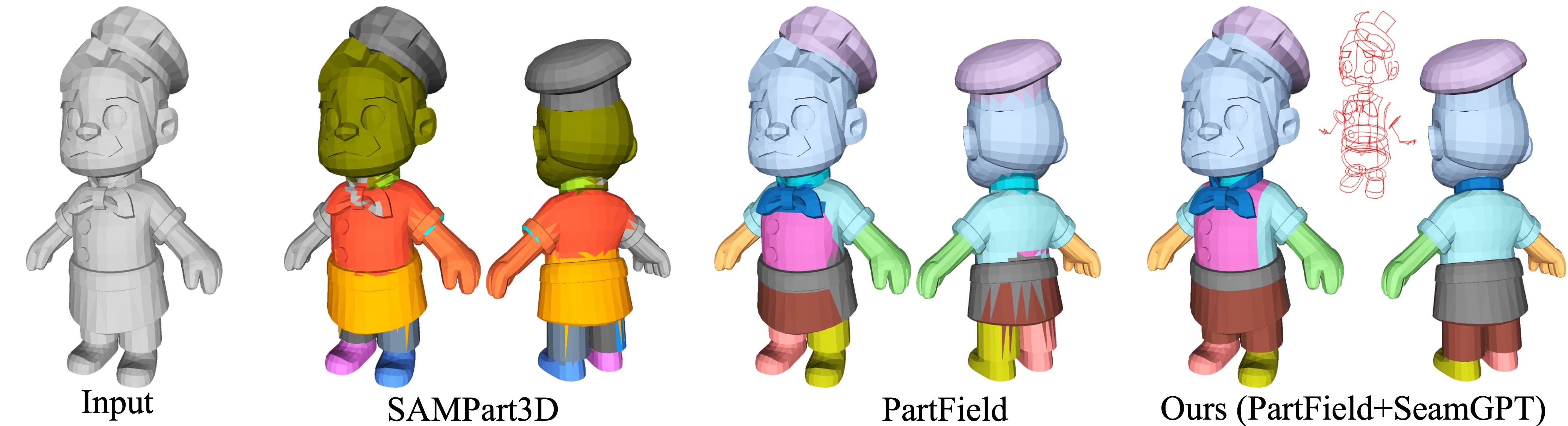}
\caption{ Qualitative part segmentation results. Our method enhances existing part segmentation methods by providing clean part boundary guidance.}
\end{figure}

\textbf{Patch-based Part Segmentation.} Given a 3D shape $S$, we first utilize PartField to obtain the initial part segmentation, resulting in a mapping of each face to a part label:
$$
L: F \to \mathcal{L}
$$
where $ F $ is the set of faces and $\mathcal{L}$ is the set of part labels. For each face $ f_i \in F $, the function $L$ maps it to a specific label $l_j \in \mathcal{L}$. Then we employ SeamGPT to predict the 3D seam lines, partitioning the shape into multiple patches $P_k$:
$$
S \to \{ P_1, P_2, \ldots, P_n \}
$$
For each patch $P_k$, we calculate the count of faces associated with different part labels:
$$
C(l_j) = \sum_{f_i \in P_k} \mathbb{I}(L(f_i) = l_j)
$$
We select the part label \( l_j^* \) with the highest count as the label for the patch $ P_k $:
$$
l_j^* = \arg\max_{l_j \in \mathcal{L}} C(l_j)
$$
Through this Patch-based Part Segmentation methodology, we can obtain well-defined parts with clean boundaries. As demonstrated in Figure~\ref{fig:segvis}, our approach significantly improves the accuracy of part decomposition, enhancing the clarity and interpretability of the segmented 3D shape.

\section{Conclusion}
\label{sec:closing}
In this paper, We present SeamGPT, an auto-regressive model that generates artist-style cutting seams for 3D meshes. By formulating surface cutting as a sequence prediction problem, our approach produces semantically meaningful and functionally coherent cuts that outperform existing methods on both UV unwrapping and part segmentation tasks. Our model effectively captures artistic cutting priors, addressing key limitations in current mesh processing workflows and offering a step toward more intelligent and semantically aware tools that better align with artist intentions. 

\textbf{Social Impact.} As a low-level building algorithm block, our work has no direct negative outcome, other than what could arise from the aforementioned applications.

\textbf{Limitations.}  Our current implementation exhibits degraded performance on meshes exceeding 20K triangular faces, requiring pre-processing via remeshing to reduce face count. 
This occurs because seam length inherently increases with mesh face count.
Potential future directions include: 1) incorporating more powerful transformer decoders to handle complex topologies and longer sequence, and 2) adopting more compact wire representations (e.g., curves) for more efficient seam handling.
Additionally, jointly learning part feature extraction and part boundary generation is a promising direction for more robust and generalizable 3D part segmentation.

\appendix

\section{Appendix}

\subsection{Metrics}
We adopt the same UV distortion metric used in FAM~\cite{zhang2024flattenanything}, which evaluates the conformality of the UV parameterization via per-face distortion. Specifically, for each triangle in the 3D mesh, we compute the Jacobian matrix \( J \in \mathbb{R}^{2 \times 3} \) that maps the 3D triangle to its flattened 2D UV counterpart via the linear relation:
\begin{equation}
    V_{2D} = J \cdot V_{3D}.
\end{equation}
Here, \( V_{3D} \in \mathbb{R}^{3 \times 3} \) and \( V_{2D} \in \mathbb{R}^{2 \times 3} \) represent the vertex positions of a triangle in 3D and 2D space, respectively. To quantify the local distortion, we compute the singular values \( \sigma_1, \sigma_2 \) of the Jacobian matrix \( J \). These values capture the local stretching or compression in the \( u \) and \( v \) directions. Ideally, conformal mappings preserve angles, corresponding to \( \sigma_1 = \sigma_2 = 1 \). Deviation from 1 reflects distortion.

Following prior works~\cite{park2021nerfies}, we compute the conformal energy per triangle as:
\begin{equation}
    E_{conf} = \left| \log \sigma_1 \right| + \left| \log \sigma_2 \right|
\end{equation}
and report the mean value across all mesh triangles. This metric effectively penalizes non-uniform scaling and captures both area and angle distortions.

\subsection{More Implemention Details}
\paragraph{Point Sampling Strategy.} Our point sampling strategy plays a critical role in guiding the transformer decoder to generate mesh-aligned seams. Instead of uniformly sampling the mesh surface, we explicitly sample points from mesh vertices and edges to ensure that the predicted seams conform to the original mesh topology.
Specifically, for each input mesh, we sample a total of 61,440 points, equally divided between vertices and edges. That is, we sample 30,720 points on vertices and 30,720 points along edges. For edge sampling, we perform uniform interpolation between edge endpoints, where the number of samples per edge is proportional to its length. If the number of vertices is fewer than 30,720, we apply repeated oversampling to reach the desired count.
This targeted sampling improves seam quality in two ways: (1) It prevents generation of seams that do not align with mesh connectivity, reducing the number of extra faces introduced during mesh cutting; and (2) it provides the transformer with more meaningful structural cues, accelerating convergence and improving generation accuracy.

\paragraph{Transformer Decoder.}
We adopt a hierarchical hourglass-style transformer decoder with causal attention, as in recent mesh autoregressive models~\cite{hao2024meshtron}. The architecture is defined by a three-level abstraction structure with depth configuration \texttt{(2, (4, 12, 4), 2)}, where each number represents the number of transformer blocks at that level. Each block has dimension 1536 and uses 16 attention heads with per-head dimension 64. Causal masking is applied at all levels to preserve autoregressive consistency.
To facilitate efficient training and decoding of long sequences, we use position encoding for a maximum sequence length of 36,864, with quantized coordinate representation at 10-bit precision (\(2^{10} = 1024\) discrete bins).

\paragraph{Sequence Control.} To regulate the length of generated seams, a length embedding is concatenated to the shape embedding. The maximum truncated length is set to 27,000 tokens, and training is performed in truncated mode to stabilize long-sequence generation.

\paragraph{Training Setup.} The model is trained with a batch size of 2, using Adam optimizer with a fixed learning rate of \(10^{-4}\), no weight decay, and gradient clipping at 0.5. A short warm-up phase of one step is used. We use data augmentation including random scaling (\(s \sim \mathcal{U}[0.95, 1.05]\)), rotation, and vertex jitter with noise level 0.01 and masking.

\subsection{More Results}

\begin{figure}
    \centering
    \includegraphics[width=1\linewidth]{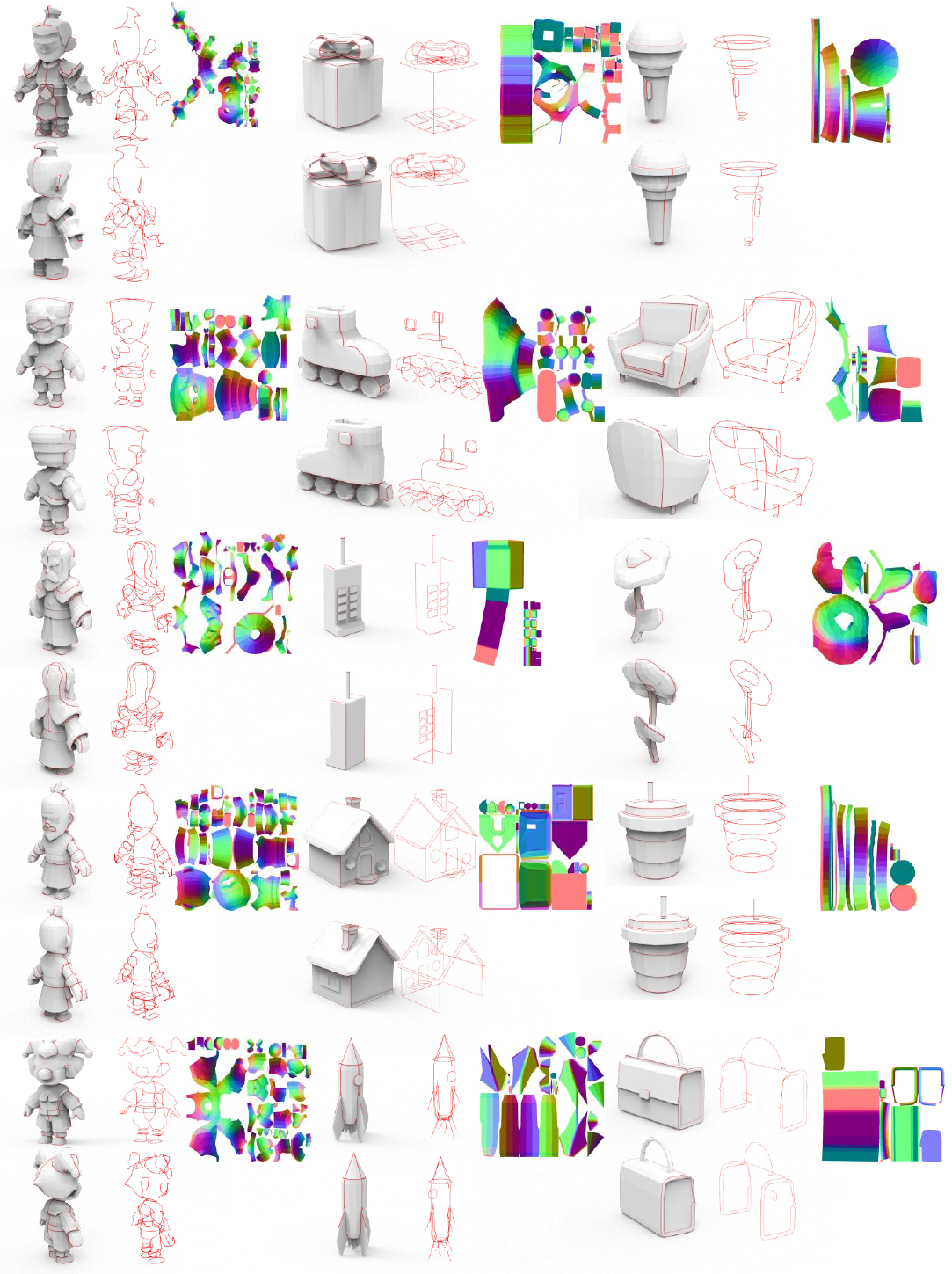}
    \caption{Seam prediction and UV unwrapping results. The left column displays the mesh with predicted seams overlaid, the middle column shows the predicted seam lines, and the right column presents the corresponding UV unwrapping results.}
    \label{fig:more_uv} 
\end{figure}

\begin{figure}
    \centering
    \includegraphics[width=1\linewidth]{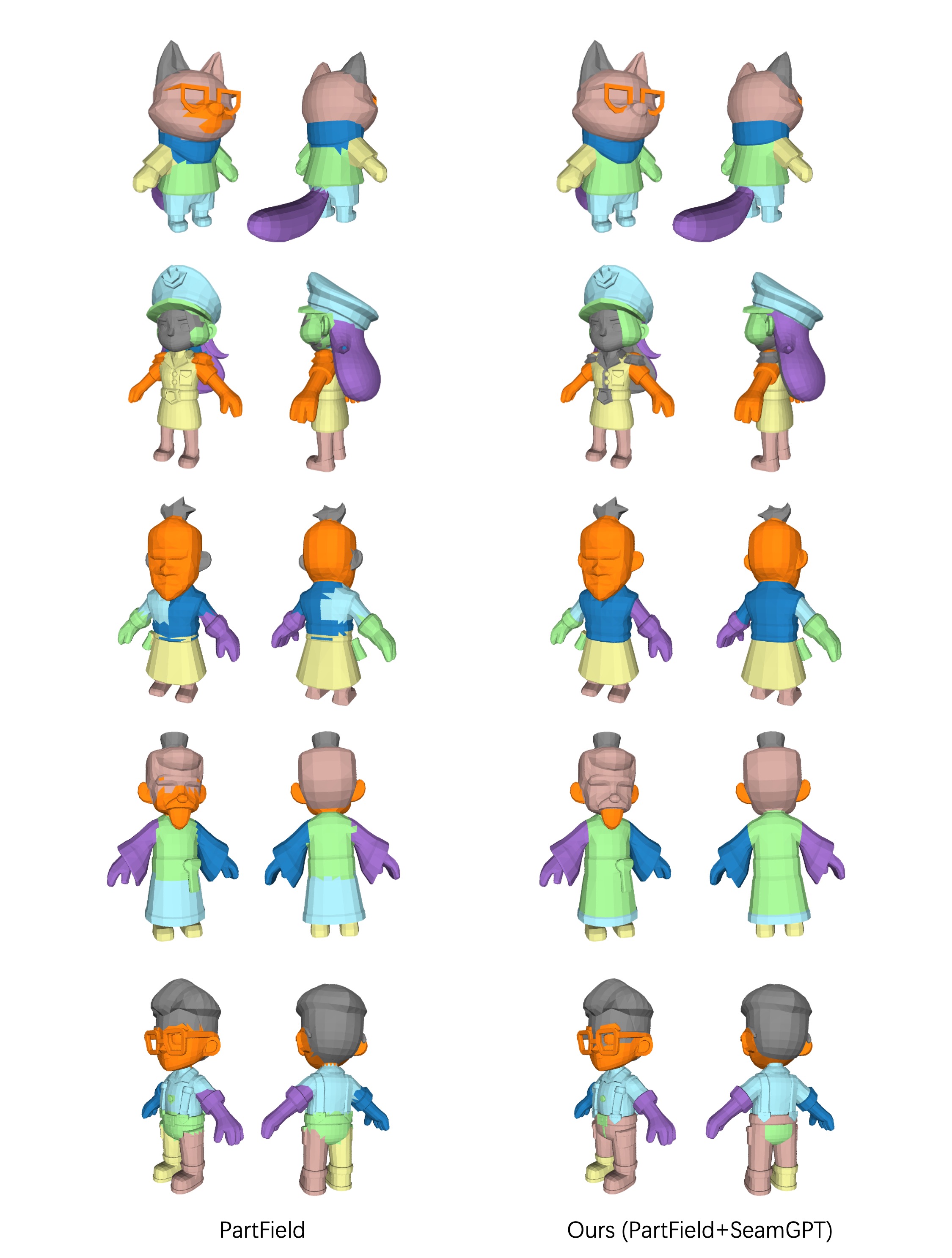}
    \caption{More part segmentation results. Last row shows the \textbf{failure case}: When SeamGPT fails to cut along the semantic boundary (e.g., between the leg and hips), this results in inconsistent segmentation with the semantics provided by PartField. }
    \label{fig:more_parts}
\end{figure}

We provide additional qualitative results of SeamGPT on diverse meshes. As shown in Figure~\ref{fig:more_uv}, our model consistently produces semantically coherent seams that align with structural boundaries.
Figure~\ref{fig:more_parts} shows additional segmentation results enhanced by SeamGPT. The predicted seams help produce cleaner and more precise part boundaries, especially in complex regions.

\bibliographystyle{abbrv}
\bibliography{main}

\end{document}